\documentclass[%
 reprint,
superscriptaddress,
 amsmath,amssymb,
 aps,
]{revtex4-2}

\usepackage{graphicx}
\usepackage{dcolumn}
\usepackage{bm}

\begin{document}

\preprint{APS/123-QED}

\title{Compton scattering study of strong orbital delocalization in LiNiO$_2$ cathode}
\author{Veenavee Nipunika Kothalawala}
\email{veenavee.kothalawala@lut.fi}
\affiliation{Department of Physics, School of Engineering Science, LUT University, FI-53851 Lappeenranta, Finland}
\author{Kosuke Suzuki}%
\affiliation{Graduate School of Science and Technology, Gunma University, Kiryu, Gunma 376-8515, Japan}
\author{Johannes Nokelainen}
\affiliation{Department of Physics, Northeastern University, Boston, Massachusetts  02115, USA}
\author{ Arttu Hyvönen}
\affiliation{Department of Physics, University of Helsinki, P.O. Box 43, FI-00014 Helsinki, Finland}
\author{Ilja Makkonen}
\affiliation{Department of Physics, University of Helsinki, P.O. Box 43, FI-00014 Helsinki, Finland}
\author{Bernardo Barbiellini}
\affiliation{Department of Physics, School of Engineering Science, LUT University, FI-53851 Lappeenranta, Finland}
\affiliation{Department of Physics, Northeastern University, Boston, Massachusetts A 02115, USA}
\author{Hasnain Hafiz}
\affiliation{Department of Physics, Northeastern University, Boston, Massachusetts  02115, USA}
\author{Pekka Tynjälä}
\affiliation{Research Unit of Sustainable Chemistry, University of Oulu, Oulu, Finland}
\affiliation{Kokkola University Consortium Chydenius, University of Jyvaskyla, Kokkola, Finland}
\author{Petteri Laine}
\affiliation{Research Unit of Sustainable Chemistry, University of Oulu, Oulu, Finland}
\affiliation{Kokkola University Consortium Chydenius, University of Jyvaskyla, Kokkola, Finland}
\author{Juho Välikangas}
\affiliation{Research Unit of Sustainable Chemistry, University of Oulu, Oulu, Finland}
\affiliation{Kokkola University Consortium Chydenius, University of Jyvaskyla, Kokkola, Finland}
\author{Tao Hu}
\affiliation{Research Unit of Sustainable Chemistry, University of Oulu, Oulu, Finland}
\author{Ulla Lassi}
\affiliation{Research Unit of Sustainable Chemistry,  University of Oulu, Oulu, Finland}
\affiliation{Kokkola University Consortium Chydenius, University of Jyvaskyla, Kokkola, Finland}
\author{Kodai Takano}%
\affiliation{Graduate School of Science and Technology, Gunma University, Kiryu, Gunma 376-8515, Japan}
\author{Naruki Tsuji}%
\affiliation{Japan Synchrotron Radiation Research Institute (JASRI), Sayo, Hyogo 679-5198, Japan}
\author{Yosuke Amada}%
\affiliation{Graduate School of Science and Technology, Gunma University, Kiryu, Gunma 376-8515, Japan}
\author{Assa Aravindh Sasikala Devi}
\affiliation{Research Unit of Sustainable Chemistry and Materials and Mechanical Engineering, University of Oulu, Pentti Kaiteran Katu 1,  FI-90570 Oulu, Finland}
\author{Matti Alatalo}
\affiliation{Research Unit of Sustainable Chemistry and Materials and Mechanical Engineering, University of Oulu,  FI-90570 Oulu, Finland}
\author{Yoshiharu Sakurai}
\affiliation{Japan Synchrotron Radiation Research Institute (JASRI), Sayo, Hyogo 679-5198, Japan}
\author{Hiroshi Sakurai}
\affiliation{Graduate School of Science and Technology, Gunma University, Kiryu, Gunma 376-8515, Japan}
\author{Arun Bansil}
\affiliation{Department of Physics, Northeastern University, Boston, Massachusetts  02115, USA}

\date{\today}

\begin{abstract}
Cobalt is used in Li-ion batteries, but it is expensive and could be replaced by nickel to deliver better performance at a lower cost.  With this motivation, we discuss how the character of redox orbitals of LiNiO$_2$ can be ascertained through x-ray Compton scattering measurements combined with parallel first-principles simulations. Our analysis reveals the 
nature of hole states in Li-doped NiO resulting from the hybridization of O 2$p$ and Ni 3$d$ orbitals. Our study also gives insight into the ferromagnetic ground state and provides a pathway toward the rational design of next-generation battery materials.

\end{abstract}

\maketitle

\section{\label{sec:level1}Introduction}
Compton scattering with hard x-rays has proven to be an effective probe of redox processes in battery materials~\cite{Hafiz2021, Nokelainen2022}.  This bulk-sensitive spectroscopy is less sensitive to defects and surfaces compared to the related positron annihilation technique \cite{pagot2023}. Both methods, however,
provide unique windows into the nature of the redox orbitals and how their subtle characteristics, such as their localization/delocalization properties, evolve during the lithiation and delithiation processes.

The first Compton scattering study 
in cathode materials
by Suzuki {\it et al.} \cite{Suzuki2015} examined the Li$_x$Mn$_2$O$_4$ (LMO) cathode and showed that the change in the shape of the Compton profile with increasing Li concentration is consistent with the appearance of states of mainly O $2p$  character. These authors also noted a negative excursion in the so-called ``difference Compton profile" (i.e., the difference between the Compton profile of the cathode with higher Li concentration and the cathode with lower Li concentration) as a basis for adducing the presence of signatures of Mn $3d$ electrons becoming more itinerant (less localized) with increasing Li content. The mechanism is that when electrons enter O $2p$ states, they hybridize with the transition metal (TM) $3d$ states and induce delocalization of the Mn $3d$ states. 
A similar study in Li$_x$CoO$_2$ (LCO)~\cite{Barbiellini2016} identified the aforementioned negative excursion as a descriptor for improved electrochemical operation in cationic redox reactions. In fact, this descriptor, called the {\em delocalization profile},
was shown to correctly point to the Li concentration range that exhibits improved cathode operation in both LMO and LCO cathodes.  
In high-capacity Li-excess systems \cite{Assat2018},
such as Li$_x$Ti$_{0.4}$Mn$_{0.4}$O$_2$ (LTMO) cationic redox reactions are complemented by anionic ones \cite{Hafiz2021}. The anionic redox reaction in LTMO produces a positive tail in the difference Compton profile spectrum at higher momentum, called the {\em Coulomb profile}, as opposed to the delocalization profile. This behavior implies that
electrons introduced into O $2p$ states repel other electrons in the Mn $3d$ states resulting in the localization of both O $2p$ and Mn $3d$ electrons.

 Interestingly, a recent study by Menon {\it et al.} \cite{menon2022} utilized high-resolution oxygen K-edge resonant inelastic x-ray spectroscopy (RIXS) to investigate LiNi$_{0.98}$W$_{0.02}$O$_2$. This study provided insight into the crucial role of oxygen ions and presented evidence of anionic redox in this non-Li-excess system. This finding is in contrast to a previous x-ray Compton scattering study by Chabaud et al. \cite{Chabaud2004} in LiNiO$_2$ (LNO) that suggested cationic redox. These contrasting results have left the question of the nature of hole states in Li-doped NiO controversial.

Here we address the oxygen character of the redox orbital in LNO and the issues raised by RIXS experiments \cite{li2019}. In this connection, utilizing high-quality LNO samples \cite{Valikangas2022a,Valikangas2022b}, we performed x-ray Compton scattering measurements along with parallel first principles modeling of the associated electronic and magnetic structures. Our analysis reveals the coexistence of orbital delocalization produced by Ni-O hybridization and the presence of a dominant O $2p$ character in the redox orbital. Our study suggests that the electronic redistribution in LNO is favorable for Li-ion battery applications and the development of Ni-rich cathode materials for next-generation Li-ion batteries \cite{Bianchini2019}.

\section{Methods}

\subsection{Samples}
Spherical Ni(OH)$_2$ precursors were synthesized using alkali metal hydroxide (NaOH) coprecipitation in an inert gas (nitrogen) atmosphere, 
see V\"alikangas {\it et al.}~\cite{Valikangas2022a,Valikangas2022b} 
and  Zhou {\it et al.}~\cite{Zhuo2010} for details. Inert gas was used to prevent the oxidation of Ni(OH)$_2$ precursor. Precipitation was conducted in a stirred-tank reactor (CSTR) with a reactor volume of 3 L at temperatures of 50°C under vigorous agitation. 
Aqueous solutions of metal sulfate NiSO$_4$·6H$_2$O, NaOH, and concentrated ammonia were fed into the reactor using peristaltic pumps. The ammonia concentration in the solution was adjusted based on the target ammonia concentration during the experiment. Particle size growth during Ni(OH)$_2$ coprecipitation was followed by determining the particle size distribution of the slurry sampled from the reactor’s overflow tubing. The solution was heated to a precipitation temperature, and the pH was adjusted to a desired level with NaOH solution. The feeding rates of nickel sulfate, NaOH, and ammonia solutions were adjusted to maintain the desired residence time.
After coprecipitation, the precursor slurry was filtered in a vacuum, and the precipitate was carefully washed with an adequate amount of deionized water. The synthesized Ni(OH)$_2$ precursors were dried overnight at 60°C in a vacuum oven. The material was milled and sieved 40 $\mu$m in dry room conditions. Ni(OH)$_2$ precursors were mixed with LiOH. Non-stoichiometric Li:Ni mole ratios from zero to one were used. Correspondingly, samples denoted LNO$_x$ were obtained for $x=0$, $x=0.25$, $x=0.5$, $x=0.75$, and $x=1$. The mixtures were calcined with a 2.5°C/min heating ramp and 5\,h holding time at a temperature of 670°C under an O$_2$ atmosphere.

The sample with a lithium concentration of $x$ can be written as Li$_x$Ni$_{2-x}$O$_2$ as confirmed by the comprehensive sample characterization process discussed in the Supplemental Material (SM). In particular, the measured x-ray diffraction patterns are in excellent agreement with theoretical predictions by Choi {\it et al.} \cite{choi2019}. Notably, Välikangas et al. (2022) have demonstrated that our fully lithiated samples exhibit high capacity and excellent capacity retention. Also, since we did not cycle the cathode materials in the current Compton study, there is no Li/Ni disordering.

\subsection{Compton profiles} 
The  Compton profile, $J(p_z)$, can be calculated
within the {\em impulse approximation} \cite{Kaplan2003, Cooper2004}
using the formula:
\begin{eqnarray}
J(p_z)= \iint \rho ({\bf p}) \,dp_{x}\,dp_{y}
\label{eq:one},
\end{eqnarray}
where {\bf p}  = ($p_x$, $p_y$, $p_z$) is the electron momentum, and $\rho ({\bf p})$ is the electron momentum density that can be expressed as:
\begin{eqnarray}
\rho({\bf p}) =  \sum_{j} n_{j}\mid\int \Psi_{j}({\bf r}) \exp(-i{\bf p \cdot r})\,
d^3{\bf r}\mid ^{2}
\label{eq:two}.
\end{eqnarray}

The electron momentum density can be cast in terms of natural orbitals $\Psi_{j}(r)$, which are the eigenstates of the one-particle density matrix of the many-electron system, and their occupation numbers, $n_{j}$ \cite{barbiellini2001}. 
Here, we consider the {\em independent particle model}, where $n_{j}=1$ if the state is occupied and $n_{j}=0$ otherwise.
The magnetic Compton profile $\it{J}_{mag}(p_z)$ is given by
\begin{equation}
J_{\mathrm{mag}}\left(p_{\mathrm{z}}\right)=\iint\left(\rho_{\uparrow}(\mathbf{p})-\rho_{\downarrow}(\mathbf{p})\right) dp_{\mathrm{x}} dp_{\mathrm{y}}
\label{eq:three}.
\end{equation}
where $\rho_{\uparrow}(\mathbf{p})$ and $\rho_{\downarrow}(\mathbf{p})$ are the momentum densities of the majority and minority spins, respectively.
The spin magnetic moment $\mu_{spin}$ is obtained by integrating $\it{J}_{mag}(p_z)$.
 We then calculate the spherical average of the Compton profile using Monte Carlo sampling coupled with linear 
interpolation. 
\\

Theoretical Compton profiles were computed
using DFT and the tabulated values
for the deep core contributions 
\cite{biggs1975}. In DFT,
the electron momentum density 
$\rho({\mathbf p})$ 
was obtained from the Kohn-Sham orbitals following Makkonen {\em et al.}~\cite{Makkonen2005} using the Vienna ab initio simulation package (VASP)~\mbox{\cite{Kresse1996, Kresse1999}} and the projector-augmented-wave (PAW) method~\cite{Blochl1994}.
The generalized gradient approximation (GGA) exchange-correlation functional parameterized by Perdew, Burke, and Ernzerhof (PBE) was used ~\cite{Perdew1996} with a kinetic energy cutoff of
520 eV for the plane-wave basis set. To account for electron correlation effects,  
a Hubbard parameter $U = 4$ eV was applied to Ni 3$d$ 
electrons following earlier studies~\cite{Tuccillo2020, Wang2006, Kothalawala2022}.  
The Brillouin zone of the supercell was 
sampled using a uniform $\Gamma$-centered 4 $\times$ 8 $\times$ 4 $k$-point grid. 

Compton and magnetic Compton profiles were measured at the high-energy inelastic scattering beamline 08W at the Japanese synchrotron facility SPring-8~\cite{Sakurai1998, Kakutani2003}. Circularly polarized x-rays of 182.6\,keV emitted from an elliptical multipole wiggler were irradiated to the sample. The size of the incident x-ray beam at the sample position is a 1\,mm square. Compton scattered x-rays were measured using a pure Ge solid-state detector. The scattering angle was fixed at 178 degrees, and Compton profiles were measured at room temperature. For magnetic Compton profiles, a magnetic field of ±2.5\,T was applied to the sample that was kept at 7\,K to obtain Compton scattered x-ray intensities, $I_+$ and $I_-$, by flipping the magnetic field every 60 s.

\section{Results and Discussion}

\begin{figure}
 \includegraphics[width=1.0\linewidth]
  {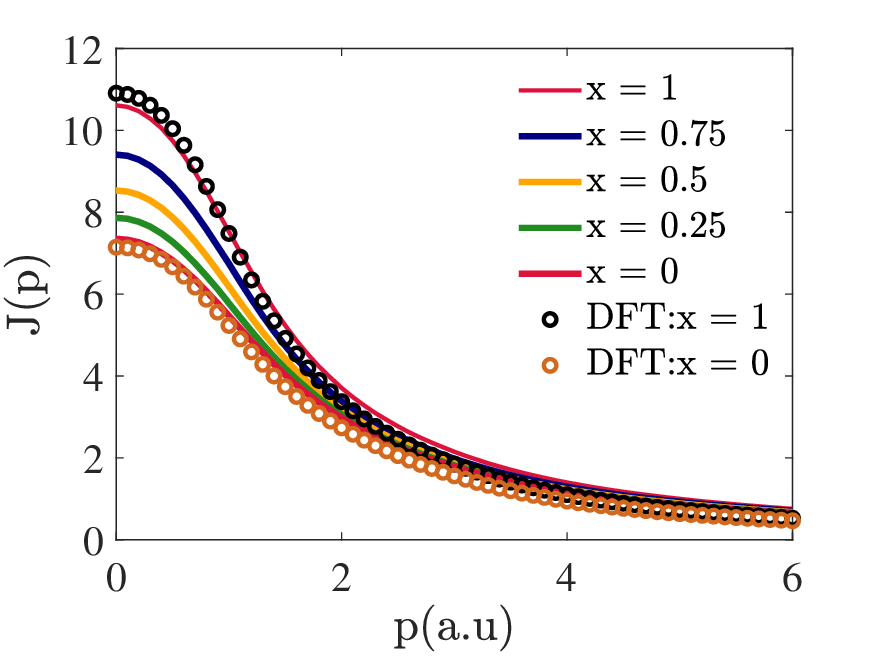} 
   \caption{Experimental and theoretical total Compton profiles (spherical averaged) for various $x$ values. Theoretical profiles are convoluted with a Gaussian of 0.5 a.u. full-width-at-half-maximum. All profiles are normalized to the electrons per nickel atom (i.e., Li$_{x/(2-x)}$NiO$_{2/(2-x)}$). 
\label{Fig:compton_profiles}}
\end{figure}

We begin by examining the Compton profiles of 
Li-doped NiO (Fig.~\ref{Fig:compton_profiles}). 
Theoretical profiles for both LNO and NiO agree well with the corresponding experimental results, indicating that the Kohn-Sham orbitals employed in our DFT calculations accurately capture the electron 
momentum distribution in Li-doped Ni oxide.

\begin{figure} 
    \centering
    \includegraphics[width=0.5\textwidth]{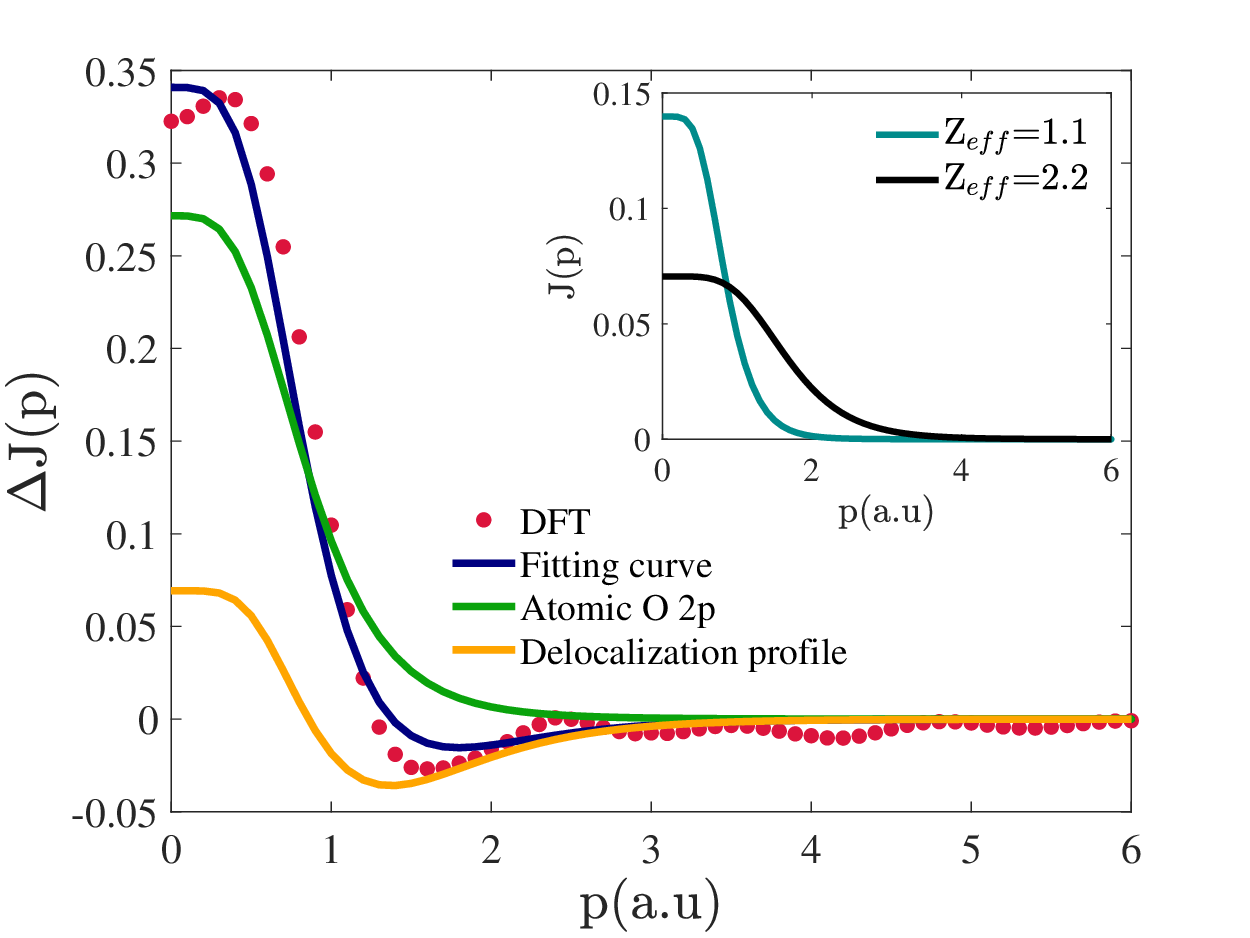}
    \caption{Theoretical difference valence Compton profiles (spherical averaged) of LNO and NiO$_2$, along with the corresponding curve-fitting results (solid lines). Contributions of O $2p$ and Ni $3d$ electrons are included. The inset provides Compton profiles of Ni 3$d$ orbital states for two different values of Zeff. The Delocaliza- tion profile is determined by the difference between the profile for Zeff = 2.2 and the profile for Zeff = 1.1. The area under the negative excursion of the profiles gives the number of   electrons delocalizing in the lithiation process. 
    \label{Fig:DFT_Fitting}}
\end{figure}

We isolate the signature of the redox orbital by subtracting the NiO$_2$ contribution from the valence Compton profile of LiNiO$_2$. The result is given by 
\begin{equation} 
\Delta J(p) = J_v^{LiNiO_2}(p) - J_v^{NiO_2}(p) \,
\label{eq:define_profile}
\end{equation}
where $J_v^{LiNiO_2}(p)$ is the spherical average valence Compton profile for LNO while $J_v^{NiO_2}(p)$ is the spherical average valence Compton profile for NiO$_2$.
Since the experimental
Compton profile of NiO$_2$ is not available; this analysis only involves our
first-principles results (FIG.~\ref{Fig:DFT_Fitting}). 
As discussed previously~\cite{Hafiz2021,Barbiellini2016}
the difference profile can be rationalized using a model with Slater orbitals, 
which involves two main contributions: the atomic O-$2p$ Compton profile and the so-called delocalization profile. To model these contributions, we utilize the radial wave functions given by Slater orbitals, characterized by effective exponents $Z_{eff}$ that are fitted to the DFT profile see SM for details.
Importance of the 
O $2p$ character has 
been previously observed in connection with XAS experiments~\cite{Kuiper1989}, while the existence of the delocalization profile $D(p)$ is necessary to explain the negative excursion 
in the difference Compton profile. The delocalization profile also allows us to quantify the number of electrons delocalizing in the lithiation process in terms of the integral $\int dp |D(p)|$.
We estimate this number to be $0.089$ electrons per Li atom.

With the preceding results in mind, we are in a position to provide a clearer explanation of the significant new physics revealed by our study regarding LNO. 
Our VASP first-principles calculations clearly show that the number of electrons in the interstitial region increases with increasing lithium concentration while the number of electrons on the nickel sites remains constant
as also shown by Genreith-Schriever and collaborators~\cite{genreith2023}. 
A similar behavior was observed by Suzuki {\em et al.} \cite{Suzuki2015} in LMO.
The number of electrons within the oxygen muffin-tin spheres also remains constant, although some of the additional charge in the interstitial region belongs to the oxygen atoms. These results strongly suggest a significant delocalization of oxygen $2p$ character and a dominant involvement of oxygen in the redox process.  This process also leads to a redistribution of Ni $3d$ electrons and results in a favorable orbital delocalization that supports battery operation.

\begin{figure}
  \includegraphics[width=1.0\linewidth]{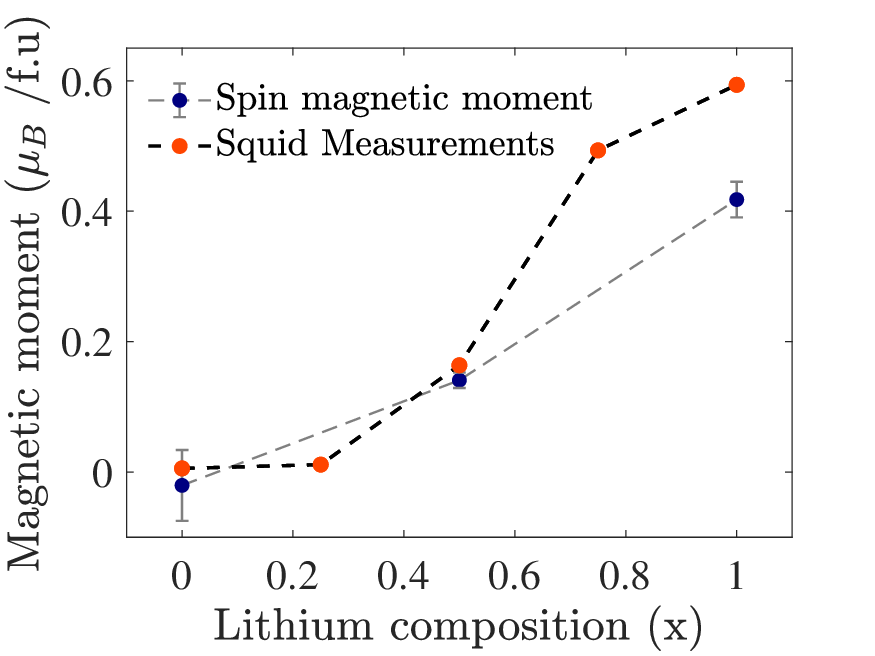} 

\caption{Spin magnetic moments for various Li concentrations $x$. The dashed line is a guide to the eye.
\label{Fig:Magnetic_moment}}
\end{figure} 

\begin{figure}
  \includegraphics[width=1.\linewidth]{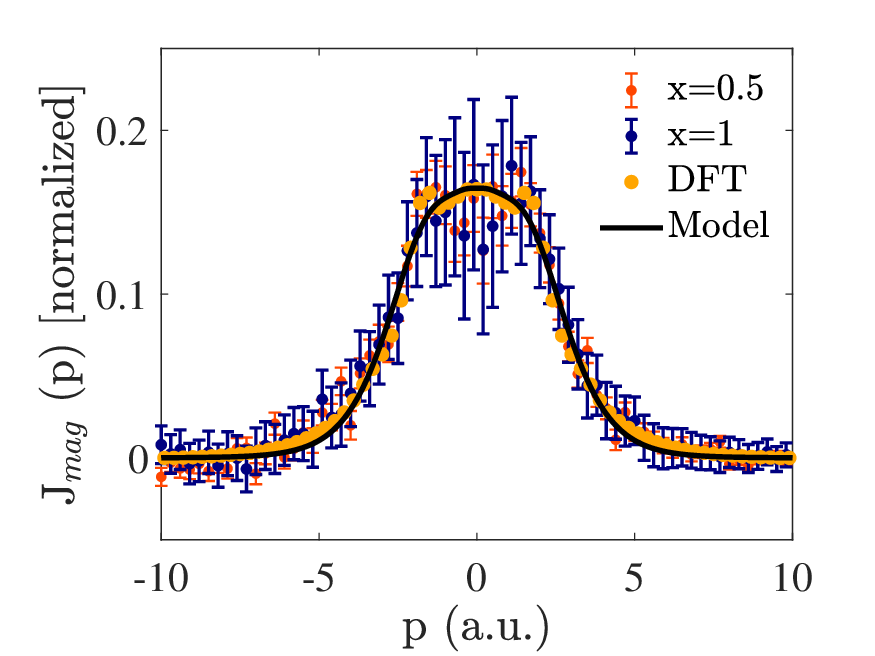}  
\caption{Magnetic Compton profiles (spherical averaged) for $x=1$, and $x=0.5$. The DFT calculation
is for $x=1$.
\label{Fig:Magnetic_profiles}}
\end{figure}

\begin{figure} 
    \centering
   \includegraphics[width=0.7\textwidth]{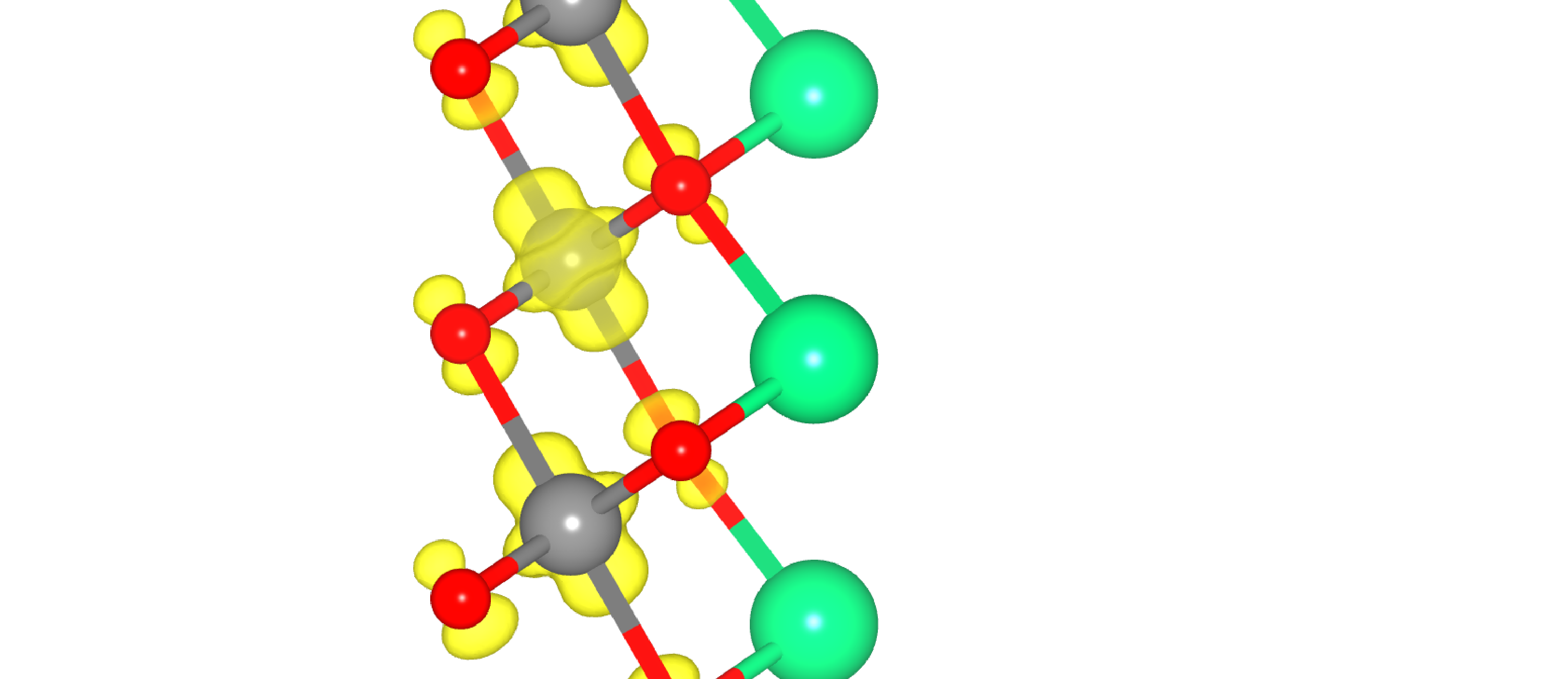}
    \caption{Spin density in LNO using an isosurface value of 0.01 a.u. on Li(green), Ni(gray), and O(red).}
    \label{Fig:Spin_density}
\end{figure}

Goodenough and colleagues first observed the ferromagnetic nature of LNO in 1958 \cite{Goodenough1958}. This intriguing ferromagnetic phase has since been confirmed 
by accurate diffusion quantum Monte Carlo (DQMC) calculations
\cite{saritas2020} and further discussed in a recent review by Tranquada \cite{Tranquada2022}. Leveraging ferromagnetism, we conducted magnetic Compton scattering experiments on LNO samples with varying Li concentrations ($x$). Figure.~\ref{Fig:Magnetic_moment} shows magnetic moments as a function of $x$ so obtained along with the values we obtained using superconducting quantum interference device (SQUID) measurements. Although the SQUID values exceed the values obtained via magnetic Compton scattering, both sets consistently yield an increasing magnetic moment with $x$. The discrepancy between the two measurements arises from the fact that SQUID measures the  total moment (i.e., orbital and spin moment), while magnetic Compton scattering only provides the spin moment.

Notably, the magnetic moment in our case did not reach saturation under the field of 2.5 T, which was used in both the magnetic Compton scattering and SQUID experiments. This lack of saturation is a significant factor contributing to the discrepancy between our measurement and our theoretical predictions, which give a magnetic moment of 1$\mu_B$ for $x=1$.

Magnetic Compton profiles for $x=1$ and  $x=0.5$ exhibit a remarkable similarity in that the two profiles can be collapsed onto a single curve when normalized to the same scale (Fig. \ref{Fig:Magnetic_profiles}). 
It is important to distinguish between the difference Compton profile and the magnetic Compton profiles, as they offer complementary insights into the redox  and magnetic orbitals, respectively \cite{Suzuki2022}.
To capture the shape of the magnetic Compton profile, we employed a fitted model using a Ni 3d Slater orbital with $Z_{eff} = 3.8$. Additionally, we included a small magnetic contribution from O 2p orbitals with $Z_{eff} = 1.1$, reflecting the mixing of a few percent due to the presence of a small local magnetic moment on the oxygen site. This fitted model is consistent with the LNO spin density calculated within the DFT, as shown in Fig. \ref{Fig:Spin_density}. Interestingly, our findings indicate a positive magnetization on the O atom, while the DQMC \cite{saritas2020} predicts a negative contribution because of the approximate 
R$\bar{3}$m crystal symmetry used in the DQMC calculation.

\section{Conclusion}
 We utilize x-ray Compton scattering along with parallel first-principles modeling to investigate the electronic structure of LNO cathode materials to gain insight into the delithiation process of going from LiNiO$_2$ to NiO$_2$. By directly measuring the electron density distribution (spherically averaged) of individual orbitals in momentum space, we showcase the unique ability of the  Compton scattering technique to extract redox and magnetic orbitals realistically. Our analysis reveals that the Ni charge in NiO, NiO$_2$, and  LiNiO$_2$ is the same, while the occupation of oxygen orbitals  changes with lithiation, indicating that O dominates the underlying charge-transfer mechanism. These results help explain oxygen loss in Ni-rich layered cathode materials, reported in recent RIXS experiments \cite{menon2022}.  Our study also sheds light on the role of $3d$ electron delocalization in the layered LNO chemistry, highlighting its importance in the redox processes.

\begin{acknowledgments}
K.S. was supported by JSPS KAKENHI Grants No. JP19K05519, No. 21KK0095, and No. 22H02103. Comp- ton and magnetic Compton scattering experiments were performed with the approval of JASRI (Proposal No. 2022A1454). Magnetization measurements were performed with the approval of the National Institute of Material Science (NIMS) Open Facility (Proposals No. 22NM8130 and No. 22NM8141). K.S. and Y.A. thank Shigeki Nimori for his support of the magnetization measurements. Part of this work at Northeastern University was supported by the Office of Naval Research Grant No. N00014-23-1-2330, and the work benefited from Northeastern University’s Advanced Scientific Computation Center and the Discovery Cluster. The authors wish to acknowledge CSC-IT Centre for Science, Finland, for computational resources.

\end{acknowledgments}

\nocite{*}
\bibliography{Reference}

\providecommand{\noopsort}[1]{}\providecommand{\singleletter}[1]{#1}%
\begin{thebibliography}{38}%
\makeatletter
\providecommand \@ifxundefined [1]{%
 \@ifx{#1\undefined}
}%
\providecommand \@ifnum [1]{%
 \ifnum #1\expandafter \@firstoftwo
 \else \expandafter \@secondoftwo
 \fi
}%
\providecommand \@ifx [1]{%
 \ifx #1\expandafter \@firstoftwo
 \else \expandafter \@secondoftwo
 \fi
}%
\providecommand \natexlab [1]{#1}%
\providecommand \enquote  [1]{``#1''}%
\providecommand \bibnamefont  [1]{#1}%
\providecommand \bibfnamefont [1]{#1}%
\providecommand \citenamefont [1]{#1}%
\providecommand \href@noop [0]{\@secondoftwo}%
\providecommand \href [0]{\begingroup \@sanitize@url \@href}%
\providecommand \@href[1]{\@@startlink{#1}\@@href}%
\providecommand \@@href[1]{\endgroup#1\@@endlink}%
\providecommand \@sanitize@url [0]{\catcode `\\12\catcode `\$12\catcode
  `\&12\catcode `\#12\catcode `\^12\catcode `\_12\catcode `\%12\relax}%
\providecommand \@@startlink[1]{}%
\providecommand \@@endlink[0]{}%
\providecommand \url  [0]{\begingroup\@sanitize@url \@url }%
\providecommand \@url [1]{\endgroup\@href {#1}{\urlprefix }}%
\providecommand \urlprefix  [0]{URL }%
\providecommand \Eprint [0]{\href }%
\providecommand \doibase [0]{https://doi.org/}%
\providecommand \selectlanguage [0]{\@gobble}%
\providecommand \bibinfo  [0]{\@secondoftwo}%
\providecommand \bibfield  [0]{\@secondoftwo}%
\providecommand \translation [1]{[#1]}%
\providecommand \BibitemOpen [0]{}%
\providecommand \bibitemStop [0]{}%
\providecommand \bibitemNoStop [0]{.\EOS\space}%
\providecommand \EOS [0]{\spacefactor3000\relax}%
\providecommand \BibitemShut  [1]{\csname bibitem#1\endcsname}%
\let\auto@bib@innerbib\@empty
\bibitem [{\citenamefont {Hafiz}\ \emph {et~al.}(2021)\citenamefont {Hafiz},
  \citenamefont {K.Suzuki}, \citenamefont {Barbiellini}, \citenamefont {Tsuji},
  \citenamefont {Yabuuchi}, \citenamefont {Yamamoto}, \citenamefont {Orikasa},
  \citenamefont {Uchimoto}, \citenamefont {Sakurai}, \citenamefont {Sakurai},
  \citenamefont {Bansil},\ and\ \citenamefont {Viswanathan}}]{Hafiz2021}%
  \BibitemOpen
  \bibfield  {author} {\bibinfo {author} {\bibfnamefont {H.}~\bibnamefont
  {Hafiz}}, \bibinfo {author} {\bibnamefont {K.Suzuki}}, \bibinfo {author}
  {\bibfnamefont {B.}~\bibnamefont {Barbiellini}}, \bibinfo {author}
  {\bibfnamefont {N.}~\bibnamefont {Tsuji}}, \bibinfo {author} {\bibfnamefont
  {N.}~\bibnamefont {Yabuuchi}}, \bibinfo {author} {\bibfnamefont
  {K.}~\bibnamefont {Yamamoto}}, \bibinfo {author} {\bibfnamefont
  {Y.}~\bibnamefont {Orikasa}}, \bibinfo {author} {\bibfnamefont
  {Y.}~\bibnamefont {Uchimoto}}, \bibinfo {author} {\bibfnamefont
  {Y.}~\bibnamefont {Sakurai}}, \bibinfo {author} {\bibfnamefont
  {H.}~\bibnamefont {Sakurai}}, \bibinfo {author} {\bibfnamefont
  {A.}~\bibnamefont {Bansil}},\ and\ \bibinfo {author} {\bibfnamefont
  {V.}~\bibnamefont {Viswanathan}},\ }\bibfield  {title} {\bibinfo {title}
  {Tomographic reconstruction of oxygen orbitals in lithium-rich battery
  materials},\ }\href {https://doi.org/10.1038/s41586-021-03509-z} {\bibfield
  {journal} {\bibinfo  {journal} {Nature}\ }\textbf {\bibinfo {volume} {564}},\
  \bibinfo {pages} {213} (\bibinfo {year} {2021})}\BibitemShut {NoStop}%
\bibitem [{\citenamefont {Nokelainen}\ \emph {et~al.}(2022)\citenamefont
  {Nokelainen}, \citenamefont {Barbiellini}, \citenamefont {Kuriplach},
  \citenamefont {Eijt}, \citenamefont {Ferragut}, \citenamefont {Li},
  \citenamefont {Kothalawala}, \citenamefont {Suzuki}, \citenamefont {Sakurai},
  \citenamefont {Hafiz}, \citenamefont {Pussi}, \citenamefont {Keshavarz},\
  and\ \citenamefont {Bansil}}]{Nokelainen2022}%
  \BibitemOpen
  \bibfield  {author} {\bibinfo {author} {\bibfnamefont {J.}~\bibnamefont
  {Nokelainen}}, \bibinfo {author} {\bibfnamefont {B.}~\bibnamefont
  {Barbiellini}}, \bibinfo {author} {\bibfnamefont {J.}~\bibnamefont
  {Kuriplach}}, \bibinfo {author} {\bibfnamefont {S.}~\bibnamefont {Eijt}},
  \bibinfo {author} {\bibfnamefont {R.}~\bibnamefont {Ferragut}}, \bibinfo
  {author} {\bibfnamefont {X.}~\bibnamefont {Li}}, \bibinfo {author}
  {\bibfnamefont {V.}~\bibnamefont {Kothalawala}}, \bibinfo {author}
  {\bibfnamefont {K.}~\bibnamefont {Suzuki}}, \bibinfo {author} {\bibfnamefont
  {H.}~\bibnamefont {Sakurai}}, \bibinfo {author} {\bibfnamefont
  {H.}~\bibnamefont {Hafiz}}, \bibinfo {author} {\bibfnamefont
  {K.}~\bibnamefont {Pussi}}, \bibinfo {author} {\bibfnamefont
  {F.}~\bibnamefont {Keshavarz}},\ and\ \bibinfo {author} {\bibfnamefont
  {A.}~\bibnamefont {Bansil}},\ }\bibfield  {title} {\bibinfo {title}
  {Identifying redox orbitals and defects in lithium-ion cathodes with compton
  scattering and positron annihilation spectroscopies: A review},\ }\bibfield
  {journal} {\bibinfo  {journal} {Condens. Matter}\ }\textbf {\bibinfo {volume}
  {7}},\ \href {https://doi.org/10.3390/condmat7030047}
  {10.3390/condmat7030047} (\bibinfo {year} {2022})\BibitemShut {NoStop}%
\bibitem [{\citenamefont {Pagot}\ \emph {et~al.}(2023)\citenamefont {Pagot},
  \citenamefont {Di~Noto}, \citenamefont {Vezz{\`u}}, \citenamefont
  {Barbiellini}, \citenamefont {Toso}, \citenamefont {Caruso}, \citenamefont
  {Zheng}, \citenamefont {Li},\ and\ \citenamefont {Ferragut}}]{pagot2023}%
  \BibitemOpen
  \bibfield  {author} {\bibinfo {author} {\bibfnamefont {G.}~\bibnamefont
  {Pagot}}, \bibinfo {author} {\bibfnamefont {V.}~\bibnamefont {Di~Noto}},
  \bibinfo {author} {\bibfnamefont {K.}~\bibnamefont {Vezz{\`u}}}, \bibinfo
  {author} {\bibfnamefont {B.}~\bibnamefont {Barbiellini}}, \bibinfo {author}
  {\bibfnamefont {V.}~\bibnamefont {Toso}}, \bibinfo {author} {\bibfnamefont
  {A.}~\bibnamefont {Caruso}}, \bibinfo {author} {\bibfnamefont
  {M.}~\bibnamefont {Zheng}}, \bibinfo {author} {\bibfnamefont
  {X.}~\bibnamefont {Li}},\ and\ \bibinfo {author} {\bibfnamefont
  {R.}~\bibnamefont {Ferragut}},\ }\bibfield  {title} {\bibinfo {title}
  {Quantum view of li-ion high mobility at carbon-coated cathode interfaces},\
  }\href@noop {} {\bibfield  {journal} {\bibinfo  {journal} {Iscience}\
  }\textbf {\bibinfo {volume} {26}},\ \bibinfo {pages} {105794} (\bibinfo
  {year} {2023})}\BibitemShut {NoStop}%
\bibitem [{\citenamefont {Suzuki}\ \emph {et~al.}(2015)\citenamefont {Suzuki},
  \citenamefont {Barbiellini}, \citenamefont {Orikasa}, \citenamefont {Go},
  \citenamefont {Sakurai}, \citenamefont {Kaprzyk}, \citenamefont {Itou},
  \citenamefont {Yamamoto}, \citenamefont {Uchimoto}, \citenamefont {Wang},
  \citenamefont {Hafiz}, \citenamefont {Bansil},\ and\ \citenamefont
  {Sakurai}}]{Suzuki2015}%
  \BibitemOpen
  \bibfield  {author} {\bibinfo {author} {\bibfnamefont {K.}~\bibnamefont
  {Suzuki}}, \bibinfo {author} {\bibfnamefont {B.}~\bibnamefont {Barbiellini}},
  \bibinfo {author} {\bibfnamefont {Y.}~\bibnamefont {Orikasa}}, \bibinfo
  {author} {\bibfnamefont {N.}~\bibnamefont {Go}}, \bibinfo {author}
  {\bibfnamefont {H.}~\bibnamefont {Sakurai}}, \bibinfo {author} {\bibfnamefont
  {S.}~\bibnamefont {Kaprzyk}}, \bibinfo {author} {\bibfnamefont
  {M.}~\bibnamefont {Itou}}, \bibinfo {author} {\bibfnamefont {K.}~\bibnamefont
  {Yamamoto}}, \bibinfo {author} {\bibfnamefont {Y.}~\bibnamefont {Uchimoto}},
  \bibinfo {author} {\bibfnamefont {Y.~J.}\ \bibnamefont {Wang}}, \bibinfo
  {author} {\bibfnamefont {H.}~\bibnamefont {Hafiz}}, \bibinfo {author}
  {\bibfnamefont {A.}~\bibnamefont {Bansil}},\ and\ \bibinfo {author}
  {\bibfnamefont {Y.}~\bibnamefont {Sakurai}},\ }\bibfield  {title} {\bibinfo
  {title} {Extracting the redox orbitals in li battery materials with
  high-resolution x-ray compton scattering spectroscopy},\ }\href
  {https://doi.org/10.1103/PhysRevLett.114.087401} {\bibfield  {journal}
  {\bibinfo  {journal} {Phys. Rev. Lett.}\ }\textbf {\bibinfo {volume} {114}},\
  \bibinfo {pages} {087401} (\bibinfo {year} {2015})}\BibitemShut {NoStop}%
\bibitem [{\citenamefont {Barbiellini}\ \emph {et~al.}(2016)\citenamefont
  {Barbiellini}, \citenamefont {Suzuki}, \citenamefont {Orikasa}, \citenamefont
  {Kaprzyk}, \citenamefont {Itou}, \citenamefont {Yamamoto}, \citenamefont
  {Wang}, \citenamefont {Hafiz}, \citenamefont {Yamada}, \citenamefont
  {Uchimoto}, \citenamefont {Bansil}, \citenamefont {Sakurai},\ and\
  \citenamefont {Sakurai}}]{Barbiellini2016}%
  \BibitemOpen
  \bibfield  {author} {\bibinfo {author} {\bibfnamefont {B.}~\bibnamefont
  {Barbiellini}}, \bibinfo {author} {\bibfnamefont {K.}~\bibnamefont {Suzuki}},
  \bibinfo {author} {\bibfnamefont {Y.}~\bibnamefont {Orikasa}}, \bibinfo
  {author} {\bibfnamefont {S.}~\bibnamefont {Kaprzyk}}, \bibinfo {author}
  {\bibfnamefont {M.}~\bibnamefont {Itou}}, \bibinfo {author} {\bibfnamefont
  {K.}~\bibnamefont {Yamamoto}}, \bibinfo {author} {\bibfnamefont {Y.~J.}\
  \bibnamefont {Wang}}, \bibinfo {author} {\bibfnamefont {H.}~\bibnamefont
  {Hafiz}}, \bibinfo {author} {\bibfnamefont {R.}~\bibnamefont {Yamada}},
  \bibinfo {author} {\bibfnamefont {Y.}~\bibnamefont {Uchimoto}}, \bibinfo
  {author} {\bibfnamefont {A.}~\bibnamefont {Bansil}}, \bibinfo {author}
  {\bibfnamefont {Y.}~\bibnamefont {Sakurai}},\ and\ \bibinfo {author}
  {\bibfnamefont {H.}~\bibnamefont {Sakurai}},\ }\bibfield  {title} {\bibinfo
  {title} {Identifying a descriptor for d-orbital delocalization in cathodes of
  li batteries based on x-ray compton scattering},\ }\bibfield  {journal}
  {\bibinfo  {journal} {Appl. Phys. Lett.}\ }\textbf {\bibinfo {volume}
  {109}},\ \href {https://doi.org/10.1063/1.4961055} {10.1063/1.4961055}
  (\bibinfo {year} {2016})\BibitemShut {NoStop}%
\bibitem [{\citenamefont {Assat}\ and\ \citenamefont
  {Tarascon}(2018)}]{Assat2018}%
  \BibitemOpen
  \bibfield  {author} {\bibinfo {author} {\bibfnamefont {G.}~\bibnamefont
  {Assat}}\ and\ \bibinfo {author} {\bibfnamefont {J.-M.}\ \bibnamefont
  {Tarascon}},\ }\bibfield  {title} {\bibinfo {title} {Fundamental
  understanding and practical challenges of anionic redox activity in li-ion
  batteries},\ }\href@noop {} {\bibfield  {journal} {\bibinfo  {journal}
  {Nature Energy}\ }\textbf {\bibinfo {volume} {3}},\ \bibinfo {pages} {373}
  (\bibinfo {year} {2018})}\BibitemShut {NoStop}%
\bibitem [{\citenamefont {Menon}\ \emph {et~al.}(2022)\citenamefont {Menon},
  \citenamefont {Johnston}, \citenamefont {Booth}, \citenamefont {Zhang},
  \citenamefont {Kress}, \citenamefont {Murdock}, \citenamefont {Fajardo},
  \citenamefont {Anthonisamy}, \citenamefont {Ruiz}, \citenamefont {Agrestini},
  \citenamefont {Zhou}, \citenamefont {Lee}, \citenamefont {Nedoma},
  \citenamefont {Cussen},\ and\ \citenamefont {Piper}}]{menon2022}%
  \BibitemOpen
  \bibfield  {author} {\bibinfo {author} {\bibfnamefont {A.~S.}\ \bibnamefont
  {Menon}}, \bibinfo {author} {\bibfnamefont {B.}~\bibnamefont {Johnston}},
  \bibinfo {author} {\bibfnamefont {S.~G.}\ \bibnamefont {Booth}}, \bibinfo
  {author} {\bibfnamefont {L.}~\bibnamefont {Zhang}}, \bibinfo {author}
  {\bibfnamefont {K.}~\bibnamefont {Kress}}, \bibinfo {author} {\bibfnamefont
  {B.}~\bibnamefont {Murdock}}, \bibinfo {author} {\bibfnamefont
  {G.}~\bibnamefont {Fajardo}}, \bibinfo {author} {\bibfnamefont {N.~N.}\
  \bibnamefont {Anthonisamy}}, \bibinfo {author} {\bibfnamefont {N.~T.}\
  \bibnamefont {Ruiz}}, \bibinfo {author} {\bibfnamefont {S.}~\bibnamefont
  {Agrestini}}, \bibinfo {author} {\bibfnamefont {K.}~\bibnamefont {Zhou}},
  \bibinfo {author} {\bibfnamefont {T.}~\bibnamefont {Lee}}, \bibinfo {author}
  {\bibfnamefont {A.}~\bibnamefont {Nedoma}}, \bibinfo {author} {\bibfnamefont
  {S.}~\bibnamefont {Cussen}},\ and\ \bibinfo {author} {\bibfnamefont
  {L.}~\bibnamefont {Piper}},\ }\bibfield  {title} {\bibinfo {title}
  {Oxygen-redox activity in non-li-excess w-doped linio$_2$ cathode},\
  }\bibfield  {journal} {\bibinfo  {journal} {ChemRxiv}\ }\href
  {https://doi.org/10.26434/chemrxiv-2022-w994j} {10.26434/chemrxiv-2022-w994j}
  (\bibinfo {year} {2022})\BibitemShut {NoStop}%
\bibitem [{\citenamefont {Chabaud}\ \emph {et~al.}(2004)\citenamefont
  {Chabaud}, \citenamefont {Bellin}, \citenamefont {Mauri}, \citenamefont
  {Loupias}, \citenamefont {Rabii}, \citenamefont {Croguennec}, \citenamefont
  {Pouillerie}, \citenamefont {Delmas},\ and\ \citenamefont
  {Buslaps}}]{Chabaud2004}%
  \BibitemOpen
  \bibfield  {author} {\bibinfo {author} {\bibfnamefont {S.}~\bibnamefont
  {Chabaud}}, \bibinfo {author} {\bibfnamefont {C.}~\bibnamefont {Bellin}},
  \bibinfo {author} {\bibfnamefont {F.}~\bibnamefont {Mauri}}, \bibinfo
  {author} {\bibfnamefont {G.}~\bibnamefont {Loupias}}, \bibinfo {author}
  {\bibfnamefont {S.}~\bibnamefont {Rabii}}, \bibinfo {author} {\bibfnamefont
  {L.}~\bibnamefont {Croguennec}}, \bibinfo {author} {\bibfnamefont
  {C.}~\bibnamefont {Pouillerie}}, \bibinfo {author} {\bibfnamefont
  {C.}~\bibnamefont {Delmas}},\ and\ \bibinfo {author} {\bibfnamefont
  {T.}~\bibnamefont {Buslaps}},\ }\bibfield  {title} {\bibinfo {title}
  {Electronic density distortion of nio$_2$ due to intercalation by li},\
  }\href {https://doi.org/https://doi.org/10.1016/j.jpcs.2003.10.021}
  {\bibfield  {journal} {\bibinfo  {journal} {J. Phys. Chem. Solids}\ }\textbf
  {\bibinfo {volume} {65}},\ \bibinfo {pages} {241} (\bibinfo {year}
  {2004})}\BibitemShut {NoStop}%
\bibitem [{\citenamefont {Li}\ \emph {et~al.}(2019)\citenamefont {Li},
  \citenamefont {Sallis}, \citenamefont {Papp}, \citenamefont {Wei},
  \citenamefont {McCloskey}, \citenamefont {Yang},\ and\ \citenamefont
  {Tong}}]{li2019}%
  \BibitemOpen
  \bibfield  {author} {\bibinfo {author} {\bibfnamefont {N.}~\bibnamefont
  {Li}}, \bibinfo {author} {\bibfnamefont {S.}~\bibnamefont {Sallis}}, \bibinfo
  {author} {\bibfnamefont {J.~K.}\ \bibnamefont {Papp}}, \bibinfo {author}
  {\bibfnamefont {J.}~\bibnamefont {Wei}}, \bibinfo {author} {\bibfnamefont
  {B.~D.}\ \bibnamefont {McCloskey}}, \bibinfo {author} {\bibfnamefont
  {W.}~\bibnamefont {Yang}},\ and\ \bibinfo {author} {\bibfnamefont
  {W.}~\bibnamefont {Tong}},\ }\bibfield  {title} {\bibinfo {title} {Unraveling
  the cationic and anionic redox reactions in a conventional layered oxide
  cathode},\ }\href {https://doi.org/10.1021/acsenergylett.9b02147} {\bibfield
  {journal} {\bibinfo  {journal} {ACS Energy Letters}\ }\textbf {\bibinfo
  {volume} {4}},\ \bibinfo {pages} {2836} (\bibinfo {year} {2019})}\BibitemShut
  {NoStop}%
\bibitem [{\citenamefont {Välikangas}\ \emph {et~al.}(2020)\citenamefont
  {Välikangas}, \citenamefont {Laine}, \citenamefont {Hietaniemi},
  \citenamefont {Hu}, \citenamefont {Tynjälä},\ and\ \citenamefont
  {Lassi}}]{Valikangas2022a}%
  \BibitemOpen
  \bibfield  {author} {\bibinfo {author} {\bibfnamefont {J.}~\bibnamefont
  {Välikangas}}, \bibinfo {author} {\bibfnamefont {P.}~\bibnamefont {Laine}},
  \bibinfo {author} {\bibfnamefont {M.}~\bibnamefont {Hietaniemi}}, \bibinfo
  {author} {\bibfnamefont {T.}~\bibnamefont {Hu}}, \bibinfo {author}
  {\bibfnamefont {P.}~\bibnamefont {Tynjälä}},\ and\ \bibinfo {author}
  {\bibfnamefont {U.}~\bibnamefont {Lassi}},\ }\bibfield  {title} {\bibinfo
  {title} {Precipitation and calcination of high-capacity linio2 cathode
  material for lithium-ion batteries},\ }\bibfield  {journal} {\bibinfo
  {journal} {Applied Sciences}\ }\textbf {\bibinfo {volume} {10}},\ \href
  {https://doi.org/10.3390/app10248988} {10.3390/app10248988} (\bibinfo {year}
  {2020})\BibitemShut {NoStop}%
\bibitem [{\citenamefont {Valikangas}\ \emph {et~al.}(2022)\citenamefont
  {Valikangas}, \citenamefont {Laine}, \citenamefont {Hietaniemi},
  \citenamefont {Hu}, \citenamefont {Selent}, \citenamefont {Tynjälä},\ and\
  \citenamefont {Lassi}}]{Valikangas2022b}%
  \BibitemOpen
  \bibfield  {author} {\bibinfo {author} {\bibfnamefont {J.}~\bibnamefont
  {Valikangas}}, \bibinfo {author} {\bibfnamefont {P.}~\bibnamefont {Laine}},
  \bibinfo {author} {\bibfnamefont {M.}~\bibnamefont {Hietaniemi}}, \bibinfo
  {author} {\bibfnamefont {T.}~\bibnamefont {Hu}}, \bibinfo {author}
  {\bibfnamefont {M.}~\bibnamefont {Selent}}, \bibinfo {author} {\bibfnamefont
  {P.}~\bibnamefont {Tynjälä}},\ and\ \bibinfo {author} {\bibfnamefont
  {U.}~\bibnamefont {Lassi}},\ }\bibfield  {title} {\bibinfo {title}
  {Correlation of aluminum doping and lithiation temperature with
  electrochemical performance of \textsc{LiNiO} cathode material},\ }\bibfield
  {journal} {\bibinfo  {journal} {J Solid State Electrochem}\ }\href
  {https://doi.org/10.1007/s10008-022-05356-y} {10.1007/s10008-022-05356-y}
  (\bibinfo {year} {2022})\BibitemShut {NoStop}%
\bibitem [{\citenamefont {Bianchini}\ \emph {et~al.}(2019)\citenamefont
  {Bianchini}, \citenamefont {Roca-Ayats}, \citenamefont {Hartmann},
  \citenamefont {Brezesinski},\ and\ \citenamefont {Janek}}]{Bianchini2019}%
  \BibitemOpen
  \bibfield  {author} {\bibinfo {author} {\bibfnamefont {M.}~\bibnamefont
  {Bianchini}}, \bibinfo {author} {\bibfnamefont {M.}~\bibnamefont
  {Roca-Ayats}}, \bibinfo {author} {\bibfnamefont {P.}~\bibnamefont
  {Hartmann}}, \bibinfo {author} {\bibfnamefont {T.}~\bibnamefont
  {Brezesinski}},\ and\ \bibinfo {author} {\bibfnamefont {J.}~\bibnamefont
  {Janek}},\ }\bibfield  {title} {\bibinfo {title} {There and back again—the
  journey of linio$_2$ as a cathode active material},\ }\href
  {https://doi.org/https://doi.org/10.1002/anie.201812472} {\bibfield
  {journal} {\bibinfo  {journal} {Angew. Chem., Int. Ed.}\ }\textbf {\bibinfo
  {volume} {58}},\ \bibinfo {pages} {10434} (\bibinfo {year}
  {2019})}\BibitemShut {NoStop}%
\bibitem [{\citenamefont {Zhou}\ \emph {et~al.}(2010)\citenamefont {Zhou},
  \citenamefont {Zhao}, \citenamefont {van A.~Bommel}, \citenamefont {Rowe},\
  and\ \citenamefont {Dahn}}]{Zhuo2010}%
  \BibitemOpen
  \bibfield  {author} {\bibinfo {author} {\bibfnamefont {F.}~\bibnamefont
  {Zhou}}, \bibinfo {author} {\bibfnamefont {X.}~\bibnamefont {Zhao}}, \bibinfo
  {author} {\bibnamefont {van A.~Bommel}}, \bibinfo {author} {\bibfnamefont
  {A.}~\bibnamefont {Rowe}},\ and\ \bibinfo {author} {\bibfnamefont {J.~R.}\
  \bibnamefont {Dahn}},\ }\bibfield  {title} {\bibinfo {title} {Coprecipitation
  synthesis of ni$_{x}$mn$_{1-x}$(oh)$_{2}$ mixed hydroxides},\ }\href
  {https://doi.org/10.1021/cm9018309} {\bibfield  {journal} {\bibinfo
  {journal} {Chemistry of Materials}\ }\textbf {\bibinfo {volume} {22}},\
  \bibinfo {pages} {1015} (\bibinfo {year} {2010})}\BibitemShut {NoStop}%
\bibitem [{\citenamefont {Choi}\ \emph {et~al.}(2019)\citenamefont {Choi},
  \citenamefont {Kang},\ and\ \citenamefont {Han}}]{choi2019}%
  \BibitemOpen
  \bibfield  {author} {\bibinfo {author} {\bibfnamefont {D.}~\bibnamefont
  {Choi}}, \bibinfo {author} {\bibfnamefont {J.}~\bibnamefont {Kang}},\ and\
  \bibinfo {author} {\bibfnamefont {B.}~\bibnamefont {Han}},\ }\bibfield
  {title} {\bibinfo {title} {Unexpectedly high energy density of a li-ion
  battery by oxygen redox in linio2 cathode: First-principles study},\
  }\href@noop {} {\bibfield  {journal} {\bibinfo  {journal} {Electrochimica
  Acta}\ }\textbf {\bibinfo {volume} {294}},\ \bibinfo {pages} {166} (\bibinfo
  {year} {2019})}\BibitemShut {NoStop}%
\bibitem [{\citenamefont {Kaplan}\ \emph {et~al.}(2003)\citenamefont {Kaplan},
  \citenamefont {Barbiellini},\ and\ \citenamefont {Bansil}}]{Kaplan2003}%
  \BibitemOpen
  \bibfield  {author} {\bibinfo {author} {\bibfnamefont {I.~G.}\ \bibnamefont
  {Kaplan}}, \bibinfo {author} {\bibfnamefont {B.}~\bibnamefont
  {Barbiellini}},\ and\ \bibinfo {author} {\bibfnamefont {A.}~\bibnamefont
  {Bansil}},\ }\bibfield  {title} {\bibinfo {title} {Compton scattering beyond
  the impulse approximation},\ }\href
  {https://doi.org/10.1103/PhysRevB.68.235104} {\bibfield  {journal} {\bibinfo
  {journal} {Phys. Rev. B}\ }\textbf {\bibinfo {volume} {68}},\ \bibinfo
  {pages} {235104} (\bibinfo {year} {2003})}\BibitemShut {NoStop}%
\bibitem [{\citenamefont {Cooper}\ \emph {et~al.}(2004)\citenamefont {Cooper},
  \citenamefont {Mijnarends}, \citenamefont {Shiorani}, \citenamefont {Sakai},\
  and\ \citenamefont {Bansil}}]{Cooper2004}%
  \BibitemOpen
  \bibfield  {author} {\bibinfo {author} {\bibfnamefont {M.}~\bibnamefont
  {Cooper}}, \bibinfo {author} {\bibfnamefont {P.}~\bibnamefont {Mijnarends}},
  \bibinfo {author} {\bibfnamefont {N.}~\bibnamefont {Shiorani}}, \bibinfo
  {author} {\bibfnamefont {N.}~\bibnamefont {Sakai}},\ and\ \bibinfo {author}
  {\bibfnamefont {A.}~\bibnamefont {Bansil}},\ }in\ \href@noop {} {\emph
  {\bibinfo {booktitle} {X-Ray Compton Scattering}}}\ (\bibinfo  {publisher}
  {Oxford University Press},\ \bibinfo {address} {Oxford},\ \bibinfo {year}
  {2004})\ pp.\ \bibinfo {pages} {31--39}\BibitemShut {NoStop}%
\bibitem [{\citenamefont {Barbiellini}\ and\ \citenamefont
  {Bansil}(2001)}]{barbiellini2001}%
  \BibitemOpen
  \bibfield  {author} {\bibinfo {author} {\bibfnamefont {B.}~\bibnamefont
  {Barbiellini}}\ and\ \bibinfo {author} {\bibfnamefont {A.}~\bibnamefont
  {Bansil}},\ }\bibfield  {title} {\bibinfo {title} {Treatment of correlation
  effects in electron momentum density: Density functional theory and beyond},\
  }\href@noop {} {\bibfield  {journal} {\bibinfo  {journal} {J. Phys. Chem.
  Solids}\ }\textbf {\bibinfo {volume} {62}},\ \bibinfo {pages} {2181}
  (\bibinfo {year} {2001})}\BibitemShut {NoStop}%
\bibitem [{\citenamefont {Biggs}\ \emph {et~al.}(1975)\citenamefont {Biggs},
  \citenamefont {Mendelsohn},\ and\ \citenamefont {Mann}}]{biggs1975}%
  \BibitemOpen
  \bibfield  {author} {\bibinfo {author} {\bibfnamefont {F.}~\bibnamefont
  {Biggs}}, \bibinfo {author} {\bibfnamefont {L.}~\bibnamefont {Mendelsohn}},\
  and\ \bibinfo {author} {\bibfnamefont {J.}~\bibnamefont {Mann}},\ }\bibfield
  {title} {\bibinfo {title} {Hartree-fock compton profiles for the elements},\
  }\href@noop {} {\bibfield  {journal} {\bibinfo  {journal} {Atomic data and
  nuclear data tables}\ }\textbf {\bibinfo {volume} {16}},\ \bibinfo {pages}
  {201} (\bibinfo {year} {1975})}\BibitemShut {NoStop}%
\bibitem [{\citenamefont {Makkonen}\ \emph {et~al.}(2005)\citenamefont
  {Makkonen}, \citenamefont {Hakala},\ and\ \citenamefont
  {Puska}}]{Makkonen2005}%
  \BibitemOpen
  \bibfield  {author} {\bibinfo {author} {\bibfnamefont {I.}~\bibnamefont
  {Makkonen}}, \bibinfo {author} {\bibfnamefont {M.}~\bibnamefont {Hakala}},\
  and\ \bibinfo {author} {\bibfnamefont {M.}~\bibnamefont {Puska}},\ }\bibfield
   {title} {\bibinfo {title} {Calculation of valence electron momentum
  densities using the projector augmented-wave method},\ }\href
  {https://doi.org/https://doi.org/10.1016/j.jpcs.2005.02.009} {\bibfield
  {journal} {\bibinfo  {journal} {Journal of Physics and Chemistry of Solids}\
  }\textbf {\bibinfo {volume} {66}},\ \bibinfo {pages} {1128} (\bibinfo {year}
  {2005})}\BibitemShut {NoStop}%
\bibitem [{\citenamefont {Kresse}\ and\ \citenamefont
  {Furthmüller}(1996)}]{Kresse1996}%
  \BibitemOpen
  \bibfield  {author} {\bibinfo {author} {\bibfnamefont {G.}~\bibnamefont
  {Kresse}}\ and\ \bibinfo {author} {\bibfnamefont {J.}~\bibnamefont
  {Furthmüller}},\ }\bibfield  {title} {\bibinfo {title} {Efficiency of
  ab-initio total energy calculations for metals and semiconductors using a
  plane-wave basis set},\ }\href {https://doi.org/10.1016/0927-0256(96)00008-0}
  {\bibfield  {journal} {\bibinfo  {journal} {Comput. Mater. Sci.}\ }\textbf
  {\bibinfo {volume} {6}},\ \bibinfo {pages} {1} (\bibinfo {year}
  {1996})}\BibitemShut {NoStop}%
\bibitem [{\citenamefont {Kresse}\ and\ \citenamefont
  {Joubert}(1999)}]{Kresse1999}%
  \BibitemOpen
  \bibfield  {author} {\bibinfo {author} {\bibfnamefont {G.}~\bibnamefont
  {Kresse}}\ and\ \bibinfo {author} {\bibfnamefont {D.}~\bibnamefont
  {Joubert}},\ }\bibfield  {title} {\bibinfo {title} {From ultrasoft
  pseudopotentials to the projector augmented-wave method},\ }\href
  {https://doi.org/10.1103/PhysRevB.59.1758} {\bibfield  {journal} {\bibinfo
  {journal} {Physical review. B}\ }\textbf {\bibinfo {volume} {59}},\ \bibinfo
  {pages} {1758} (\bibinfo {year} {1999})}\BibitemShut {NoStop}%
\bibitem [{\citenamefont {Bl\"ochl}(1994)}]{Blochl1994}%
  \BibitemOpen
  \bibfield  {author} {\bibinfo {author} {\bibfnamefont {P.~E.}\ \bibnamefont
  {Bl\"ochl}},\ }\bibfield  {title} {\bibinfo {title} {Projector augmented-wave
  method},\ }\href {https://doi.org/10.1103/PhysRevB.50.17953} {\bibfield
  {journal} {\bibinfo  {journal} {Phys. Rev. B}\ }\textbf {\bibinfo {volume}
  {50}},\ \bibinfo {pages} {17953} (\bibinfo {year} {1994})}\BibitemShut
  {NoStop}%
\bibitem [{\citenamefont {Perdew}\ \emph {et~al.}(1996)\citenamefont {Perdew},
  \citenamefont {Burke},\ and\ \citenamefont {Ernzerhof}}]{Perdew1996}%
  \BibitemOpen
  \bibfield  {author} {\bibinfo {author} {\bibfnamefont {J.}~\bibnamefont
  {Perdew}}, \bibinfo {author} {\bibfnamefont {K.}~\bibnamefont {Burke}},\ and\
  \bibinfo {author} {\bibfnamefont {M.}~\bibnamefont {Ernzerhof}},\ }\bibfield
  {title} {\bibinfo {title} {Generalized gradient approximation made simple},\
  }\href {https://doi.org/10.1103/PhysRevLett.77.3865} {\bibfield  {journal}
  {\bibinfo  {journal} {Phys. Rev. Lett.}\ }\textbf {\bibinfo {volume} {77}},\
  \bibinfo {pages} {3865} (\bibinfo {year} {1996})}\BibitemShut {NoStop}%
\bibitem [{\citenamefont {Tuccillo}\ \emph {et~al.}(2000)\citenamefont
  {Tuccillo}, \citenamefont {Palumbo}, \citenamefont {Pavone}, \citenamefont
  {Muñoz-García}, \citenamefont {Paolone},\ and\ \citenamefont
  {Brutti}}]{Tuccillo2020}%
  \BibitemOpen
  \bibfield  {author} {\bibinfo {author} {\bibfnamefont {M.}~\bibnamefont
  {Tuccillo}}, \bibinfo {author} {\bibfnamefont {O.}~\bibnamefont {Palumbo}},
  \bibinfo {author} {\bibfnamefont {M.}~\bibnamefont {Pavone}}, \bibinfo
  {author} {\bibfnamefont {A.}~\bibnamefont {Muñoz-García}}, \bibinfo
  {author} {\bibfnamefont {A.}~\bibnamefont {Paolone}},\ and\ \bibinfo {author}
  {\bibfnamefont {S.}~\bibnamefont {Brutti}},\ }\bibfield  {title} {\bibinfo
  {title} {Analysis of the phase stability of limo$_2$ layered oxides (m = co,
  mn, ni)},\ }\href {https://doi.org/10.3390/cryst10060526} {\bibfield
  {journal} {\bibinfo  {journal} {Crystals}\ }\textbf {\bibinfo {volume}
  {10}},\ \bibinfo {pages} {526} (\bibinfo {year} {2000})}\BibitemShut
  {NoStop}%
\bibitem [{\citenamefont {Wang}\ \emph {et~al.}(2006)\citenamefont {Wang},
  \citenamefont {Maxisch},\ and\ \citenamefont {Ceder}}]{Wang2006}%
  \BibitemOpen
  \bibfield  {author} {\bibinfo {author} {\bibfnamefont {L.}~\bibnamefont
  {Wang}}, \bibinfo {author} {\bibfnamefont {T.}~\bibnamefont {Maxisch}},\ and\
  \bibinfo {author} {\bibfnamefont {G.}~\bibnamefont {Ceder}},\ }\bibfield
  {title} {\bibinfo {title} {Analysis of the phase stability of limo$_2$
  layered oxides (m = co, mn, ni)},\ }\href
  {https://doi.org/10.1103/PhysRevB.73.195107} {\bibfield  {journal} {\bibinfo
  {journal} {Phys. Rev. B}\ }\textbf {\bibinfo {volume} {73}},\ \bibinfo
  {pages} {195107} (\bibinfo {year} {2006})}\BibitemShut {NoStop}%
\bibitem [{\citenamefont {Kothalawala}\ \emph {et~al.}(2022)\citenamefont
  {Kothalawala}, \citenamefont {Devi}, \citenamefont {Nokelainen},
  \citenamefont {Alatalo}, \citenamefont {Barbiellini}, \citenamefont {Hu},
  \citenamefont {Lassi}, \citenamefont {Suzuki}, \citenamefont {Sakurai},\ and\
  \citenamefont {Bansil}}]{Kothalawala2022}%
  \BibitemOpen
  \bibfield  {author} {\bibinfo {author} {\bibfnamefont {V.~N.}\ \bibnamefont
  {Kothalawala}}, \bibinfo {author} {\bibfnamefont {A.~A.~S.}\ \bibnamefont
  {Devi}}, \bibinfo {author} {\bibfnamefont {J.}~\bibnamefont {Nokelainen}},
  \bibinfo {author} {\bibfnamefont {M.}~\bibnamefont {Alatalo}}, \bibinfo
  {author} {\bibfnamefont {B.}~\bibnamefont {Barbiellini}}, \bibinfo {author}
  {\bibfnamefont {T.}~\bibnamefont {Hu}}, \bibinfo {author} {\bibfnamefont
  {U.}~\bibnamefont {Lassi}}, \bibinfo {author} {\bibfnamefont
  {K.}~\bibnamefont {Suzuki}}, \bibinfo {author} {\bibfnamefont
  {H.}~\bibnamefont {Sakurai}},\ and\ \bibinfo {author} {\bibfnamefont
  {A.}~\bibnamefont {Bansil}},\ }\bibfield  {title} {\bibinfo {title} {First
  principles calculations of the optical response of linio$_{2}$},\ }\bibfield
  {journal} {\bibinfo  {journal} {Condens. Matter}\ }\textbf {\bibinfo {volume}
  {7}},\ \href {https://doi.org/10.3390/condmat7040054}
  {10.3390/condmat7040054} (\bibinfo {year} {2022})\BibitemShut {NoStop}%
\bibitem [{\citenamefont {Sakurai}(1998)}]{Sakurai1998}%
  \BibitemOpen
  \bibfield  {author} {\bibinfo {author} {\bibfnamefont {Y.}~\bibnamefont
  {Sakurai}},\ }\bibfield  {title} {\bibinfo {title} {{High-Energy
  Inelastic-Scattering Beamline for Electron Momentum Density Study}},\ }\href
  {https://doi.org/10.1107/S0909049598002052} {\bibfield  {journal} {\bibinfo
  {journal} {Journal of Synchrotron Radiation}\ }\textbf {\bibinfo {volume}
  {5}},\ \bibinfo {pages} {208} (\bibinfo {year} {1998})}\BibitemShut {NoStop}%
\bibitem [{\citenamefont {Kakutani}\ \emph {et~al.}(2003)\citenamefont
  {Kakutani}, \citenamefont {Kubo}, \citenamefont {Koizumi}, \citenamefont
  {Sakai}, \citenamefont {Ahuja},\ and\ \citenamefont {Sharma}}]{Kakutani2003}%
  \BibitemOpen
  \bibfield  {author} {\bibinfo {author} {\bibfnamefont {Y.}~\bibnamefont
  {Kakutani}}, \bibinfo {author} {\bibfnamefont {Y.}~\bibnamefont {Kubo}},
  \bibinfo {author} {\bibfnamefont {A.}~\bibnamefont {Koizumi}}, \bibinfo
  {author} {\bibfnamefont {N.}~\bibnamefont {Sakai}}, \bibinfo {author}
  {\bibfnamefont {B.~L.}\ \bibnamefont {Ahuja}},\ and\ \bibinfo {author}
  {\bibfnamefont {B.~K.}\ \bibnamefont {Sharma}},\ }\bibfield  {title}
  {\bibinfo {title} {Magnetic compton profiles of fcc-ni, fcc-fe50ni50 and
  hcp-co},\ }\href {https://doi.org/10.1143/JPSJ.72.599} {\bibfield  {journal}
  {\bibinfo  {journal} {Journal of the Physical Society of Japan}\ }\textbf
  {\bibinfo {volume} {72}},\ \bibinfo {pages} {599} (\bibinfo {year}
  {2003})}\BibitemShut {NoStop}%
\bibitem [{\citenamefont {Kuiper}\ \emph {et~al.}(1989)\citenamefont {Kuiper},
  \citenamefont {Kruizinga}, \citenamefont {Ghijsen}, \citenamefont
  {Sawatzky},\ and\ \citenamefont {Verweij}}]{Kuiper1989}%
  \BibitemOpen
  \bibfield  {author} {\bibinfo {author} {\bibfnamefont {P.}~\bibnamefont
  {Kuiper}}, \bibinfo {author} {\bibfnamefont {G.}~\bibnamefont {Kruizinga}},
  \bibinfo {author} {\bibfnamefont {J.}~\bibnamefont {Ghijsen}}, \bibinfo
  {author} {\bibfnamefont {G.~A.}\ \bibnamefont {Sawatzky}},\ and\ \bibinfo
  {author} {\bibfnamefont {H.}~\bibnamefont {Verweij}},\ }\bibfield  {title}
  {\bibinfo {title} {Character of holes in
  ${\mathrm{li}}_{x}{\mathrm{ni}}_{1\ensuremath{-}x}\mathrm{O}$ and their
  magnetic behavior},\ }\href {https://doi.org/10.1103/PhysRevLett.62.221}
  {\bibfield  {journal} {\bibinfo  {journal} {Phys. Rev. Lett.}\ }\textbf
  {\bibinfo {volume} {62}},\ \bibinfo {pages} {221} (\bibinfo {year}
  {1989})}\BibitemShut {NoStop}%
\bibitem [{\citenamefont {Genreith-Schriever}\ \emph
  {et~al.}(2023)\citenamefont {Genreith-Schriever}, \citenamefont {Banerjee},
  \citenamefont {Menon}, \citenamefont {Bassey}, \citenamefont {Piper},
  \citenamefont {Grey},\ and\ \citenamefont {Morris}}]{genreith2023}%
  \BibitemOpen
  \bibfield  {author} {\bibinfo {author} {\bibfnamefont {A.~R.}\ \bibnamefont
  {Genreith-Schriever}}, \bibinfo {author} {\bibfnamefont {H.}~\bibnamefont
  {Banerjee}}, \bibinfo {author} {\bibfnamefont {A.~S.}\ \bibnamefont {Menon}},
  \bibinfo {author} {\bibfnamefont {E.~N.}\ \bibnamefont {Bassey}}, \bibinfo
  {author} {\bibfnamefont {L.~F.}\ \bibnamefont {Piper}}, \bibinfo {author}
  {\bibfnamefont {C.~P.}\ \bibnamefont {Grey}},\ and\ \bibinfo {author}
  {\bibfnamefont {A.~J.}\ \bibnamefont {Morris}},\ }\bibfield  {title}
  {\bibinfo {title} {Oxygen hole formation controls stability in linio2
  cathodes},\ }\href@noop {} {\bibfield  {journal} {\bibinfo  {journal}
  {Joule}\ }\textbf {\bibinfo {volume} {7}},\ \bibinfo {pages} {1623} (\bibinfo
  {year} {2023})}\BibitemShut {NoStop}%
\bibitem [{\citenamefont {Goodenough}\ \emph {et~al.}(1958)\citenamefont
  {Goodenough}, \citenamefont {Wickham},\ and\ \citenamefont
  {Croft}}]{Goodenough1958}%
  \BibitemOpen
  \bibfield  {author} {\bibinfo {author} {\bibfnamefont {J.~B.}\ \bibnamefont
  {Goodenough}}, \bibinfo {author} {\bibfnamefont {D.~G.}\ \bibnamefont
  {Wickham}},\ and\ \bibinfo {author} {\bibfnamefont {W.~J.}\ \bibnamefont
  {Croft}},\ }\bibfield  {title} {\bibinfo {title} {Some ferrimagnetic
  properties of the system
  ${\mathrm{li}}_{x}{\mathrm{ni}}_{1\ensuremath{-}x}\mathrm{O}$},\ }\href
  {https://doi.org/10.1063/1.1723144} {\bibfield  {journal} {\bibinfo
  {journal} {J. Appl. Phys.}\ }\textbf {\bibinfo {volume} {29}},\ \bibinfo
  {pages} {382} (\bibinfo {year} {1958})}\BibitemShut {NoStop}%
\bibitem [{\citenamefont {Saritas}\ \emph {et~al.}(2020)\citenamefont
  {Saritas}, \citenamefont {Fadel}, \citenamefont {Kozinsky},\ and\
  \citenamefont {Grossman}}]{saritas2020}%
  \BibitemOpen
  \bibfield  {author} {\bibinfo {author} {\bibfnamefont {K.}~\bibnamefont
  {Saritas}}, \bibinfo {author} {\bibfnamefont {E.~R.}\ \bibnamefont {Fadel}},
  \bibinfo {author} {\bibfnamefont {B.}~\bibnamefont {Kozinsky}},\ and\
  \bibinfo {author} {\bibfnamefont {J.~C.}\ \bibnamefont {Grossman}},\
  }\bibfield  {title} {\bibinfo {title} {Charge density and redox potential of
  linio2 using ab initio diffusion quantum monte carlo},\ }\href@noop {}
  {\bibfield  {journal} {\bibinfo  {journal} {The Journal of Physical Chemistry
  C}\ }\textbf {\bibinfo {volume} {124}},\ \bibinfo {pages} {5893} (\bibinfo
  {year} {2020})}\BibitemShut {NoStop}%
\bibitem [{\citenamefont {Tranquada}(2022)}]{Tranquada2022}%
  \BibitemOpen
  \bibfield  {author} {\bibinfo {author} {\bibfnamefont {J.~M.}\ \bibnamefont
  {Tranquada}},\ }\bibfield  {title} {\bibinfo {title} {John goodenough and the
  many lives of transition-metal oxides},\ }\bibfield  {journal} {\bibinfo
  {journal} {J. Electrochem. Soc.}\ }\textbf {\bibinfo {volume} {169}},\ \href
  {https://doi.org/10.1149/1945-7111/ac4895} {10.1149/1945-7111/ac4895}
  (\bibinfo {year} {2022})\BibitemShut {NoStop}%
\bibitem [{\citenamefont {Suzuki}\ \emph {et~al.}(2022)\citenamefont {Suzuki},
  \citenamefont {Hoshi}, \citenamefont {Sakurai}, \citenamefont {Tsuji},
  \citenamefont {Yamamoto}, \citenamefont {Yabuuchi}, \citenamefont {Hafiz},
  \citenamefont {Orikasa}, \citenamefont {Uchimoto}, \citenamefont {Sakurai},
  \citenamefont {Viswanathan}, \citenamefont {Bansil},\ and\ \citenamefont
  {Barbiellini}}]{Suzuki2022}%
  \BibitemOpen
  \bibfield  {author} {\bibinfo {author} {\bibfnamefont {K.}~\bibnamefont
  {Suzuki}}, \bibinfo {author} {\bibfnamefont {Y.~O.~K.}\ \bibnamefont
  {Hoshi}}, \bibinfo {author} {\bibfnamefont {H.}~\bibnamefont {Sakurai}},
  \bibinfo {author} {\bibfnamefont {N.}~\bibnamefont {Tsuji}}, \bibinfo
  {author} {\bibfnamefont {K.}~\bibnamefont {Yamamoto}}, \bibinfo {author}
  {\bibfnamefont {N.}~\bibnamefont {Yabuuchi}}, \bibinfo {author}
  {\bibfnamefont {H.}~\bibnamefont {Hafiz}}, \bibinfo {author} {\bibfnamefont
  {Y.}~\bibnamefont {Orikasa}}, \bibinfo {author} {\bibfnamefont
  {Y.}~\bibnamefont {Uchimoto}}, \bibinfo {author} {\bibfnamefont
  {Y.}~\bibnamefont {Sakurai}}, \bibinfo {author} {\bibfnamefont
  {V.}~\bibnamefont {Viswanathan}}, \bibinfo {author} {\bibfnamefont
  {A.}~\bibnamefont {Bansil}},\ and\ \bibinfo {author} {\bibfnamefont
  {B.}~\bibnamefont {Barbiellini}},\ }\bibfield  {title} {\bibinfo {title}
  {Magnetic compton scattering study of li-rich battery materials},\ }\bibfield
   {journal} {\bibinfo  {journal} {Condensed Matter}\ }\textbf {\bibinfo
  {volume} {7}},\ \href {https://doi.org/10.3390/condmat7010004}
  {10.3390/condmat7010004} (\bibinfo {year} {2022})\BibitemShut {NoStop}%
\bibitem [{FES()}]{FESEM}%
  \BibitemOpen
  \href@noop {} {\bibinfo {title} {Fesem, carl zeiss microscopy gmbh,
  germany}}\BibitemShut {NoStop}%
\bibitem [{Rig()}]{Rigaku}%
  \BibitemOpen
  \href@noop {} {\bibinfo {title} {A rigaku smartlab}}\BibitemShut {NoStop}%
\bibitem [{The()}]{ThermoFisher}%
  \BibitemOpen
  \href@noop {} {\bibinfo {title} {Thermo electron icap 6500 duo, thermofisher
  scientific, usa}}\BibitemShut {NoStop}%
\bibitem [{Qua()}]{Quantum}%
  \BibitemOpen
  \href@noop {} {\bibinfo {title} {Quantum design japan,inc}}\BibitemShut
  {NoStop}%
\end{thebibliography}%
\end{document}